\documentclass[preprint,pra,showpacs,preprintnumbers,amsmath,amssymb]{revtex4-1}

\usepackage[utf8]{inputenc}
\usepackage{amsmath, amssymb,graphicx}
\usepackage[cdot,mediumqspace,squaren]{SIunits}
\usepackage{braket}
\usepackage{mathrsfs}

\newcommand{\E}{\vec{E}} 
\newcommand{\Ham}{\hat{H}} 
\newcommand{\cAnni}{\hat{a}} 
\newcommand{\cCreat}{\hat{a}^\dagger} 
\renewcommand{\vec}{\mathbf} 
\newcommand{\e}[1]{\text{e}^{#1}} 
\renewcommand{\k}{\vec{k}} 
\renewcommand{\Im}[1]{\text{Im}\left\{#1\right\}} 
\newcommand{\eps}{ \varepsilon} 
\newcommand{\db}[1]{\underline{#1}} 
\newcommand{\p}{\vec{p}} 
\newcommand{\op}[1]{\hat{#1}} 

\graphicspath{{./images/},{./}}

\begin{document}

\title{Negative spontaneous emission by a moving two-level atom}

\author{Sylvain Lanneb\`{e}re\textsuperscript{1}}
\author{M\'{a}rio G. Silveirinha\textsuperscript{1,2}}
\email{To whom correspondence should be addressed:
mario.silveirinha@co.it.pt} \affiliation{\textsuperscript{1}
Department of Electrical Engineering, University of Coimbra and
Instituto de Telecomunica\c{c}\~{o}es, 3030-290 Coimbra, Portugal}
\affiliation{\textsuperscript{2}University of Lisbon -- Instituto
Superior T\'ecnico, Department of Electrical Engineering, 1049-001
Lisboa, Portugal}

\date{\today}

\begin{abstract}
In this paper we investigate how the dynamics of a two-level atom is
affected by its interaction with the quantized near field of a
plasmonic slab in relative motion. We demonstrate that for small
separation distances and a relative velocity greater than a certain
threshold, this interaction can lead to a population inversion, such
that the probability of the excited state exceeds the probability of
the ground state, corresponding to a negative spontaneous emission
rate. It is shown that the developed theory is intimately related to
a classical problem. The problem of quantum friction is analyzed and
the differences with respect to the corresponding classical effect
are highlighted.
\end{abstract}

\date{\today}

\maketitle

\section{Introduction}

Since the pioneering experimental work of Drexhage
\cite{drexhage_monomolecular_1970, drexhage_iv_1974}, it is well
known that an excited atom placed in close vicinity of a planar
metallic surface sees its spontaneous emission rate strongly
affected by the presence of the surface. This phenomenon plays a
role particularly important in  modern nano optics and is partly due
to the coupling between the atom and the Surface Plasmon Polaritons
(SPPs) supported by the metallic surface
\cite{chance_molecular_1978, novotny_principles_2012}. In this
paper, we are interested in a related scenario and investigate how
the spontaneous emission rate is modified when the metallic surface
and the two-level atom are in relative translational motion.

The electromagnetic interactions between polarizable moving matter
have been extensively studied in the literature, notably in the
context of the dynamical Casimir effect
\cite{fulling_radiation_1976,schwinger_casimir_1992,barton_quantum_1993,barton_quantum_1996,eberlein_sonoluminescence_1996,dodonov_current_2010,maslovski_friction_2013,maghrebi_scattering_2013,
silveirinha_theory_2014, Henkel_2015}. In particular, it was
recently proven that two closely separated materials in relative
translational motion may start to spontaneously emit light due to
the emergence of optical instabilities
\cite{silveirinha_theory_2014,silveirinha_quantization_2013,
maslovski_friction_2013}, an effect related to the Vavilov-Cherenkov
radiation but for neutral matter \cite{meyer_quantum_1985}. This
phenomenon occurs because of the coupling between the guided modes
supported by the moving bodies, and results from the conversion of
kinetic energy into electromagnetic energy, originating a quantum
friction effect
\cite{teodorovich_contribution_1978,levitov_van_1989,mkrtchian_interaction_1995,pendry_shearing_1997,volokitin_theory_1999,volokitin_near-field_2007,scheel_casimir-polder_2009,philbin_no_2009,pendry_quantum_2010,leonhardt_comment_2010,pendry_reply_2010,barton_van_2010,dedkov_tangential_2012,horshley_canonical_2012,Hoye_Casimir_2013,pieplow_fully_2013,volokitin_comment_2014,Hoye_Casimir_2014,silveirinha_theory_2014,Milton_Reality_2015,Hoye_Casimir_2015,hoye_casimir_friction_2015}.
From the quantum mechanical point of view, it can be understood as a
consequence of the existence of oscillators associated with negative
frequencies (due to the Doppler shift effect) that behave as energy
reservoirs, and thereby serve to pump the wave oscillations and
generate the unstable behavior
\cite{silveirinha_quantization_2013,silveirinha_optical_2014,
horshley_negative_2016, guo_singular_2014,lannebere_wave_2016}. Interestingly, similar
optical instabilities were predicted to appear for a classical
electric dipole moving in the vicinity of a metallic surface
supporting surface-plasmon polaritons
\cite{silveirinha_optical_2014}. The objective of this article is to
look at the same problem from the point of view of quantum
electrodynamics using the Markov approximation and modeling the
point dipole as a two-level atom. It should be mentioned that the
friction force and spontaneous emission by a neutral atom
interacting with the near-field of a dispersive dielectric in
relative motion was studied by different authors (see a review in
Refs. \cite{Milton_Reality_2015, Henkel_2015}). In particular, in
Refs. \cite{horshley_canonical_2012,horshley_negative_2016} it was
shown that a quantum harmonic oscillator initially in its ground
state can be excited by the quantized field of a moving dielectric
slab, in qualitative agreement with our general findings. Here, we
consider instead that the dipole is a two-level atom and that the
slab has a plasmonic response, and connect the negative spontaneous
emission rate with the unstable response of the corresponding
classical problem.

The paper is organized as follows. In Sec. \ref{sec:formalism} we
describe the geometry of the problem and the adopted formalism.
Then, in Sec. \ref{sec:Fermi_golden_rule} the spontaneous emission
rate experienced by the moving atom is determined using the Fermi's
golden rule. It is demonstrated that the relative motion may induce
non-conservative transitions, which eventually lead to a negative
rate of spontaneous emission. In Sec. \ref{sec:self_field} we use
classical electrodynamics to find the decay rate of the natural
oscillations of a point dipole, and show its relation with the
quantum spontaneous emission rate. Then, in section
\ref{sec:evol_atom_op}, the time evolution of the atomic operators
expectation is characterized, and it is shown that when the rate of
spontaneous emission becomes negative, the system evolves towards an
inversion of population. Moreover, it is verified that a quantum
friction force emerges in the stationary regime. Finally, in order
to illustrate the concepts developed throughout the paper, Sec.
\ref{sec:Numerical_results} is devoted to numerical examples in the
quasi-static approximation and in particular the conditions to have
a negative spontaneous emission are given. The friction force
dependence on the atom velocity is also characterized.


\section{The system under-study} \label{sec:formalism}

\begin{figure}[!ht]
\centering
\includegraphics[width=.35\linewidth]{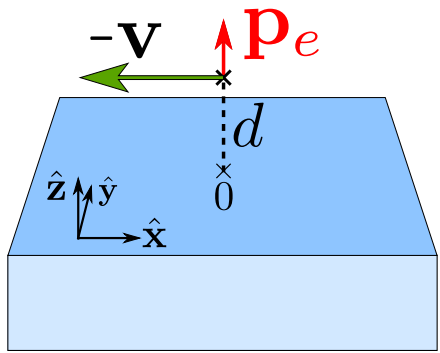}
         \caption{The system under study: a two-level atom is placed at a distance $d$ from a thick metallic surface.
         The two-level atom moves with a relative velocity $-\vec{v}$ with respect to the metal.}
\label{fig:system_under_study}
\end{figure}

The system studied here is closely related to that considered in
Ref. \cite{silveirinha_optical_2014} and consists of an electric
dipole (modeled as a two-level atom in the quantum case) located at
a distance $d$ from a thick metallic surface of infinite extent
along the $x$ and $y$ directions, as depicted in Fig.
\ref{fig:system_under_study}.

It is assumed that the permittivity of the metallic slab is
described by a lossless Drude model. The relative velocity of the
metal slab with respect to the two-level atom is $\vec{v}=v
\hat{\vec{x}}$. Hence, in the frame of the metal slab the two-level
atom has velocity  $-\vec{v}$. It is supposed that the velocity is
time-independent, and hence the effect of the optically induced
friction-type force -- which acts to reduce the relative velocity
between the atom and metal -- is neglected
\cite{silveirinha_optical_2014, Milton_Reality_2015, Henkel_2015}.
It is assumed here and throughout the paper that the unprimed
coordinates refer to the reference system co-moving with the slab,
and the primed coordinates to the reference system co-moving with
the atom. It is also supposed that the relative velocity is
significantly smaller than the speed of light in vacuum so that the
spacetime coordinates in the two frames can be linked by a Galilean
transformation. Moreover, in harmonic regime we adopt the time
variation $\e{-i \omega t}$ for the electromagnetic fields.

\subsection{Quantized electromagnetic field}

As is well known, the quantized electromagnetic field is obtained by
associating to each normal mode of the classical problem a quantum
harmonic oscillator. For the system of interest, the electromagnetic
modes satisfy:
\begin{align} \label{E:Maxwell_equation_freq}
  \op{N} \vec{F} = \omega \vec{M} \cdot \vec{F},
\end{align}
where $\vec{F}=\left(\vec{E} \quad \vec{H} \right)^T$ stands for a
six-vector whose components are the electric and the magnetic
fields. In the above, $\op{N}$ is a differential operator
\begin{align}
\op{N}= \begin{pmatrix} 0 & i \nabla \times \vec{1}_{3\times3} \\ -i
\nabla \times \vec{1}_{3\times3} & 0 \end{pmatrix},
\end{align}
and $\vec{M}$ is the material matrix
\begin{align}
\vec{M}(\vec{r},\omega)=\begin{pmatrix} \eps(\vec{r},\omega)
\vec{1}_{3\times3} & \bf{0} \\ \bf{0} & \mu_0 \vec{1}_{3\times3},
\end{pmatrix}
\end{align}
being $\eps$ the space-dependent dispersive permittivity of the
system. The relevant materials are assumed non-magnetic. In our
problem the system is invariant to translations along the $x$ and
$y$ directions and hence the transverse wave vector ${\bf{k}} =
\left( {{k_x},{k_y},0} \right)$ determines two good quantum numbers.
Hence, the electromagnetic modes are of the form
$\vec{F}_{n\k}(\vec{r})=\vec{f}_{n\k} (z) \e{i \k \cdot \vec{r}}$
where $\vec{f}_{n\k}$ is the field envelope which depends only on
the $z$ coordinate. Then, the quantized fields in the frame
co-moving with the plasmonic slab can be written as
\cite{knoll_1987, glauber_1991, Sipe_2006,
silveirinha_exchange_2012}:
\begin{align}\label{E:quantized_field}
\op{\vec{F}}(\vec{r},t) = \sum_{\omega_{n\k}>0} \sqrt{\frac{\hbar
\omega_{n\k} }{2}} ( \cAnni_{n\k}(t) \vec{F}_{n\k}(\vec{r})  +
\cCreat_{n\k}(t) \vec{F}_{n\k}^\ast(\vec{r})  ),
\end{align}
where the sum is restricted to positive oscillation frequencies
$\omega_{n\k}$ and $\cAnni_{n\k}$ and $\cCreat_{n\k}$ are the photon
annihilation and creation operators for the mode ${n\k}$, which obey
the commutation relation $\left[
\cAnni_{n\k},\cCreat_{m\vec{q}}\right]=\delta_{n,m}\delta_{\vec{q},\k}$.
In these conditions the field Hamiltonian is
\begin{equation}
 \Ham_\text{field}=  \sum_{\omega_{n\k}>0} \frac{\hbar \omega_{n\k}}{2} \left(\cCreat_{n\k} \cAnni_{n\k} +  \cAnni_{n\k}\cCreat_{n\k}   \right).
\end{equation}
For dispersive materials the electromagnetic modes must be
normalized as:
\begin{equation} \label{E:relation_normalization_modes}
 \braket{\vec{F}_{n\vec{k}}|\vec{F}_{n\k}}= \frac{1}{2} \int d^3\vec{r}~ \vec{F}_{n\vec{k}}^\ast \cdot \frac{\partial \left[\omega\vec{M}\right]}{\partial \omega} \cdot \vec{F}_{n\k} =1.
\end{equation}
This normalization condition is consistent with Ref.
\cite{Sipe_2006}. We note in passing that the time-dependent Maxwell
equations in dispersive media can always be transformed into a
generalized system with no dispersion using the formalism developed
in Ref. \cite{silveirinha_chern_2015} (see also Appendix
\ref{sec:appendix_Green's_function}). In this context, the material
dispersion is described by additional variables representing the
internal degrees of freedom of the material
\cite{raman_photonic_2010,morgado_analytical_2015,silveirinha_chern_2015}.
Interestingly, the transformed problem can be readily quantized
using standard techniques \cite{ silveirinha_exchange_2012} because
the corresponding spectral problem is a standard linear Hermitian
eigenvalue problem. In particular, in this framework the
normalization condition \eqref{E:relation_normalization_modes}
emerges naturally from Eq. (10) of Ref.
\cite{silveirinha_chern_2015}.

\subsection{Interaction with the two-level atom}
In the quantum description the electric dipole is modeled as a two-level
atom whose excited and ground states $\ket{e}$ and $\ket{g}$
are separated by the energy $E_e-E_g=\hbar\omega_0$, where
$\omega_0$ is the atomic transition frequency. By choosing the
energy origin at $(E_e+E_g)/2$, the Hamiltonian of the atom can
be written as
\begin{align}
 \Ham_\text{at}&= \frac{\hbar}{2} \omega_0 \op{\sigma}_z,
\end{align}
with $\op{\sigma}_z=\ket{e}\bra{e} - \ket{g}\bra{g}$ the atomic
inversion operator.

For low field intensities and a wavelength far greater than the Bohr
radius, the Hamiltonian representing the interaction between the two-level
atom and the quantized field can be expressed in the  dipole
approximation as $\Ham_\text{int}=-\op{\vec{p}}_e\cdot
\hat{\vec{E}}'(\vec{r}_0')$ \cite{haroche_exploring_2006}, where
$\hat{\vec{E}}'(\vec{r}_0')$ is the electric field evaluated at the
two-level atom position $(\vec{r}_0')$ in the atom co-moving frame, and
$\op{\vec{p}}_e$ is the electric dipole moment operator given by
$\op{\vec{p}}_e=-e \op{\vec{r}}$. If the states $\ket{e}$ and
$\ket{g}$ have opposite parities, then the odd operator
$\op{\vec{r}}$ has only non-diagonal components in the
$(\ket{e},\ket{g})$ basis and the dipole moment is therefore of the
form $\op{\vec{p}}_e= \boldsymbol{\gamma}_{eg}^\ast \op{\sigma}_+ +
\boldsymbol{\gamma}_{eg} \op{\sigma}_-$. Here, $\op{\sigma}_+ =
\ket{e}\bra{g}$ describes a transition to the excited state,
$\op{\sigma}_- = \ket{g}\bra{e}$ describes a transition to the
ground state, and the vector $\boldsymbol{\gamma}_{eg}$ represents
the transition electric dipole moment.

Throughout this paper the non-relativistic regime is assumed so that
the electromagnetic fields in the two reference frames are simply
related by $\op{\vec{F}}'(\vec{r}_0',t')\approx
\op{\vec{F}}(\vec{r}_0,t )$ with $\vec{r}_0=\vec{r}_0'-\vec{v} t$
\cite{jackson_classical_1999}.
Then, with these approximations and using the six-vector formalism
the interaction Hamiltonian can be written as
\begin{equation}
  \Ham_\text{int} \approx - (\tilde{\boldsymbol{\gamma}}_{eg}^\ast  \op{\sigma}_+ + \tilde{\boldsymbol{\gamma}}_{eg} \op{\sigma}_-) \cdot   \hat{\vec{F}}(\vec{r}_0'-\vec{v} t),
\end{equation}
where $\tilde{\boldsymbol{\gamma}}_{eg}=\left(
\boldsymbol{\gamma}_{eg} \quad 0 \right)^T$ is a six-vector. Using
the expression of the quantized field \eqref{E:quantized_field} and
taking into account the dependence of classical fields in the
transverse ($x$ and $y$) coordinates it follows that:
\begin{align}\label{E:interaction_Hamiltonian}
  \Ham_\text{int}  \approx  - ( \tilde{\boldsymbol{\gamma}}_{eg}^\ast \op{\sigma}_+  + \tilde{\boldsymbol{\gamma}}_{eg} \op{\sigma}_-)  \cdot \sum_{\omega_{n\k}>0}
  \sqrt{\frac{\hbar \omega_{n\k} }{2}} ( \cAnni_{n\k} \vec{F}_{n\k}(\vec{r}_0') \e{-i \k \cdot \vec{v} t }  + \cCreat_{n\k} \vec{F}_{n\k}^\ast(\vec{r}_0') \e{i \k \cdot \vec{v} t } ),
\end{align}
It is important to note that this Hamiltonian induces four types of
transitions. Two transitions are associated with the so-called
energy conserving terms (proportional to $\op{\sigma}_+
\cAnni_{n\k}$ or $\op{\sigma}_- \cCreat_{n\k}$), and the other two
transitions describe non-conservative processes that increase
($\op{\sigma}_+ \cCreat_{n\k}$) or decrease ($\op{\sigma}_-
\cAnni_{n\k}$) both quantum numbers. As shown in the following
sections, the non-conservative processes -- which play no role when
the relative velocity vanishes -- can become dominant for large
velocities and small separation distances.

\section{Quantum spontaneous emission rate} \label{sec:Fermi_golden_rule}

Next, the Fermi's golden rule is used to obtain the spontaneous
emission rate experienced by the two-level atom when moving in the
vicinity of the plasmonic slab. According to the Fermi's golden rule
\cite{shankar_principles_1994}, the number of transitions per unit
time $R_{\ket{i} \to \ket{f}}$ from an initial state $\ket{i}$ with
energy $E_i = \hbar \omega_i$ to a final state $\ket{f}$ with energy
$E_f = \hbar \omega_f$ can be expressed as
\begin{equation} \label{E:Fermi_golde_rate}
R_{\ket{i} \to \ket{f}}=\lim_{t \to \infty}  \frac{ \left| \int_0^t
\bra{f} \Ham_\text{int} \ket{i} \e{i (\omega_f-\omega_i) t'} ~dt'
\right|^2 }{ \hbar^2 t},
\end{equation}
where the states are assumed to belong to a continuum. We are
interested in two competing processes. The first process is the
usual spontaneous emission transition between the excited state of
the atom to an excited state of the field with a single photon:
${\ket{e,0} \to \ket{g,1_{n\k}} }$ (here the second index refers to
the field state). Using the interaction Hamiltonian
\eqref{E:interaction_Hamiltonian} in Eq. \eqref{E:Fermi_golde_rate}
and summing over all possible radiation channels it is found that
the transition rate for this process is
\begin{equation}
 \Gamma^+ =    \sum_{\omega_{n\k}>0}    \frac{\pi}{\hbar}    \omega_{n\k} \left|
 \tilde{\boldsymbol{\gamma}}_{eg}
 \cdot  \vec{F}_{n\k}^\ast(\vec{r}_0' ) \right|^2    \delta\left( \omega_0 - \omega_{n\k}' \right) , \label{E:gammaPlus}
\end{equation}
where $\omega_{n\k}'=\omega_{n\k} + \k \cdot \vec{v}$ is the Doppler
shifted frequency in the reference frame of the two-level atom.
Evidently, the described mechanism drives the atom to its ground
state.

The second process of relevance corresponds to a transition between
the ground state of the atom to an excited state of the field:
$\ket{g,0} \to \ket{e,1_{n\k}}$. \emph{A priori} this process may
seem impossible because $\ket{g,0}$ is the ground state of the
non-interacting atom and field, but a simple calculation shows that
its time rate is determined by:
\begin{equation}
 \Gamma^- =    \sum_{\omega_{n\k}>0}    \frac{\pi}{\hbar}    \omega_{n\k} \left|
 \tilde{\boldsymbol{\gamma}}_{eg}^\ast
  \cdot  \vec{F}_{n\k}^\ast(\vec{r}_0' ) \right|^2 \delta\left( \omega_0 + \omega_{n\k}' \right)   .\label{E:gammaMoins}
 \end{equation}
Evidently, this process drives the atom (and the field) to an
excited state, even though there is no photon absorption. When the
relative velocity between the atom and the metal vanishes,
$\Gamma^-=0$, as expected. However, when there is relative motion
the Doppler shifted frequencies $\omega_{n\k}'$ may be negative, and
hence the relative motion may induce transitions to the excited
state.

In general, transitions between ${\ket{e,0} \to \ket{g,1_\k} }$
involve \textit{only positive} Doppler shifted frequencies such that
$\omega_{n\k}'=\omega_0$. These transitions are generated by the
energy conserving terms in Eq. \eqref{E:interaction_Hamiltonian}. On
the other hand, transitions between $\ket{g,0} \to \ket{e,1_\k} $
involve \textit{only negative} Doppler shifted frequencies such that
$\omega_{n\k}'=-\omega_0$, and thus it follows that they can only
occur when the relative velocity between the slab and the atom is
large enough to provide $\omega_{n\k}'<0$. Such transitions are
induced by the non-conservative terms in the interaction Hamiltonian
\eqref{E:interaction_Hamiltonian}, and the conservation of the total
energy in the system implies that they should be associated with a
reduction of the kinetic energy
\cite{nezlin_negative-energy_1976,frolov_excitation_1986,silveirinha_optical_2014,silveirinha_theory_2014}.

We define the total spontaneous emission rate $\Gamma_\text{gr}$
from the excited state to the ground, $\ket{e} \to \ket{g}$, as the
difference between the transition rates of the two competing
processes:
\begin{align} \label{E:Gamma_golden_rule}
\Gamma_\text{gr} &= \Gamma^+ - \Gamma^-  \nonumber\\
&=\frac{\pi }{\hbar }\sum\limits_{{\omega _{n{\bf{k}}}} > 0}
{{\omega _{n{\bf{k}}}}} \left[ {{{\left| {{\bf{\tilde \gamma }}_{eg}
\cdot {\bf{F}}_{n{\bf{k}}}^*\left( \vec{r}_0' \right)}
\right|}^2}\delta \left( { \omega_{n\k}' - {\omega _0}} \right) -
{{\left| {{{{\bf{\tilde \gamma }}}_{eg}}^* \cdot
{\bf{F}}_{n{\bf{k}}}^*\left( \vec{r}_0' \right)} \right|}^2}\delta
\left( {  \omega_{n\k}' + {\omega _0}} \right)} \right].
\end{align}
The physical meaning of $\Gamma_\text{gr}$ will be further
elaborated later. Of course, in a situation with no relative motion
the contribution of $\Gamma^-$ vanishes and the spontaneous emission
rate $\Gamma_\text{gr}$ reduces to the usual result for a dipole
located near a stationary metallic surface
\cite{novotny_principles_2012}. Notably, it will be shown ahead that
it is possible to find a regime in which $\Gamma^->\Gamma^+$ i.e. a
situation where the interaction with the negative Doppler shifted
frequencies is dominant,  such that the total spontaneous emission
rate becomes negative, implying that the state $\ket{g,0}$ becomes
less likely of being occupied than the state $\ket{e,0}$.

\section{The decay rate for the classical problem} \label{sec:self_field}

In order to link the spontaneous emission rate $\Gamma_\text{gr}$
with a classical result, we note that a neutral two-level atom can be
modeled classically as an electric dipole (with dipole moment
$\vec{p}_e$) with a polarizability (along the relevant direction of
space) given by the semi-classical formula
\cite{loudon_theory_2006,silveirinha_optical_2014}
\begin{equation}
 \alpha_e^{-1}(\omega)= \frac{\hbar \omega_0 \eps_0}{|\gamma_{eg}|^2} \left( \frac{1}{2\omega_0^2} (\Gamma_0^2+\omega_0^2-\omega^2)-i\Gamma_0 \frac{\omega}{\omega_0^2}\right),
\end{equation}
where $\omega_0$ is the transition frequency of the two-level atom,
$\gamma_{eg}$ is the transition dipole moment and
$\Gamma_0=\frac{|\gamma_{eg}|^2}{6\pi\eps_0\hbar} \left(
\frac{\omega_0}{c} \right)^3$. When the electric dipole is in the
vicinity of the plasmonic slab, the natural frequencies of
oscillation, $\omega=\omega' + i \omega''$, are determined by the
solutions of
\begin{equation} \label{E:charac_eq_classical}
\boldsymbol{\gamma}_{eg}^\ast \cdot \left[\alpha_e^{-1}(\omega) -
\db{C}_\text{int}(\omega) \right] \cdot \boldsymbol{\gamma}_{eg} =0,
\end{equation}
where $\db{C}_\text{int}$ is an interaction dyadic that relates the
local field acting on the dipole with the electric dipole moment:
${{\bf{E}}'_{{\rm{loc}}}} = \db{C}_\text{int} \cdot
{{\bf{p}}_e}/{\varepsilon _0}$ \cite{silveirinha_optical_2014}. The
local field is the total field excluding the self-field radiated by
the dipole, and depends on the metal slab and on the relative
velocity. The resonances of atomic systems are characterized by
large quality factors, and hence the solution of
\eqref{E:charac_eq_classical} can be obtained with perturbation
theory. The solution is of the form $\omega \approx \omega_0 + i
\omega''$ with the imaginary part given by
\cite{silveirinha_optical_2014}:
\begin{equation}
   \omega''\approx -\frac{1}{\hbar  \eps_0} \Im{ \boldsymbol{\gamma}_{eg}^\ast \cdot \left[ i \frac{1}{6\pi} \left( \frac{\omega_0}{c} \right)^3 + \db{C}_\text{int}(\omega_0) \right] \cdot \boldsymbol{\gamma}_{eg}}.
\end{equation}
It is convenient to rewrite this result as:
\begin{equation}
   \omega''\approx -\frac{1}{\hbar  \eps_0} \Im{ \boldsymbol{\gamma}_{eg}^\ast \cdot \db{C}_\text{tot}(\omega_0) \cdot \boldsymbol{\gamma}_{eg}}.
\end{equation}
where $\db{C}_\text{tot}$ is the interaction dyadic that relates the
total electric field (including the self-field of the dipole) with
the electric dipole moment: ${\bf{E}}' = \db{C}_\text{tot} \cdot
{{\bf{p}}_e}/{\varepsilon _0}$. Noting that the electromagnetic
energy is proportional to $\e{2 \omega''t}$, it follows that the
classical decay rate of the dipole is $\Gamma_\text{cl}=-2\omega''$,
or more explicitly
\begin{equation} \label{E:gamma_classical}
   \Gamma_\text{cl}\approx \frac{2}{\hbar  \eps_0} \Im{ \boldsymbol{\gamma}_{eg}^\ast \cdot \db{C}_\text{tot}(\omega_0) \cdot \boldsymbol{\gamma}_{eg}}.
\end{equation}
As expected, in the absence of interactions with the metallic slab
($\db{C}_\text{int}=0$), the classical decay rate reduces to the
free space spontaneous emission's rate: $\Gamma_\text{sp} = \frac{
|\gamma_{eg}|^2}{3\pi \hbar  \eps_0} \left( \frac{\omega_0}{c}
\right)^3
 $. Next, we demonstrate that this result is generally valid and
 that $\Gamma_\text{cl}$ is coincident with the quantum spontaneous emission
 rate $\Gamma_\text{gr}$ determined in the previous section [Eq. \eqref{E:Gamma_golden_rule}].

In Appendix  \ref{sec:appendix_field_dipole}, the field radiated by
the moving classical dipole is explicitly calculated based on an
eigenmode expansion. Using a non-relativistic approximation, it is
shown that in the reference frame co-moving with the dipole the
field is determined by Eq. \eqref{E:F_dip_final}. Let us introduce
an $6 \times 6$ interaction dyadic $\db{C}_\text{tot}^\text{g}$
defined in such a manner that:
\begin{equation}  \label{E:definition_generalized_Cint}
\vec{F}'(\vec{r}_0')= \db{C}_\text{tot}^\text{g}(\omega) \cdot
\vec{p},
\end{equation}
where $\p=\left( \p_e \quad 0 \right)^T$ is the generalized dipole
moment (a six-vector). From Eq. \eqref{E:F_dip_final}, it is clear
that:
\begin{equation}\label{E:Cint_eigenmodes}
\db{C}_\text{tot}^\text{g}(\omega) =   \sum_{\omega_{n\k}>0}
\frac{\omega_{n\k}}{2 } \left( \frac{1 }{ \omega_{n\k}' -\omega}
\vec{F}_{n\k}(\vec{r}_0') \otimes  \vec{F}^\ast_{n\k}(\vec{r}_0') +
\frac{1 }{ \omega_{n\k}' +\omega }  \vec{F}^\ast_{n\k}(\vec{r}_0')
\otimes  \vec{F}_{n\k}(\vec{r}_0')  \right)   - \vec{M}_0^{-1}
\delta(0).
\end{equation}
where $\vec{M}_0$ is the material matrix in the free-space region
and the positive frequency eigenmodes are normalized as in
Eq.\eqref{E:relation_normalization_modes}. Note that
$\db{C}_\text{tot} / \varepsilon_0$ is the electric sub-component of
$\db{C}_\text{tot}^\text{g}$ (the $3\times3$ sub-block matrix in the
upper-left corner). Using $\frac{1}{\omega_n \pm \omega}=
\mathcal{P}\frac{1}{\omega_n \pm \omega} \mp i  \pi \delta(\omega_n
\pm \omega)$ (where $\mathcal{P}$ stands for the principal value
operator), it follows that the anti-Hermitian part of the
interaction dyadic is
\begin{align} \label{E:Im_Cint_eigenmodes}
\frac{1}{{2i}}
&\left[\db{C}_\text{tot}^\text{g}(\omega)-\db{C}_\text{tot}^\text{g}(\omega)^\dag\right]
= \nonumber\\
&\sum_{\omega_{n\k}>0} \frac{\pi \omega_{n\k} }{2 } \left[
\delta(\omega_{n\k}' -\omega) \vec{F}_{n\k}(\vec{r}_0') \otimes
\vec{F}^\ast_{n\k}(\vec{r}_0') - \delta(\omega_{n\k}' +\omega)
\vec{F}^\ast_{n\k}(\vec{r}_0') \otimes \vec{F}_{n\k}(\vec{r}_0')
\right].
\end{align}
Hence, by direct comparison with the quantum emission rate obtained
with the Fermi's golden rule \eqref{E:Gamma_golden_rule} and
recalling that $\tilde{\boldsymbol{\gamma}}_{eg}=\left(
\boldsymbol{\gamma}_{eg} \quad 0 \right)^T$, it is seen that:
\begin{align} \label{E:Gamma_Cint_golden_rule}
\Gamma_\text{gr} = \frac{2 }{\hbar }
   \Im{ \tilde{\boldsymbol{\gamma}}_{eg}^\ast \cdot
\db{C}_\text{tot}^\text{g}(\omega_0) \cdot
\tilde{\boldsymbol{\gamma}}_{eg}} .
\end{align}
We used the property ${\mathop{\rm Im}\nolimits} \left\{
{{\bf{w}}{}^* \cdot {\bf{A}} \cdot {\bf{w}}} \right\} = {\bf{w}}{}^*
\cdot \frac{{{\bf{A}} - {{\bf{A}}^\dag }}}{{2i}} \cdot {\bf{w}}$. It
is evident that the right-hand side of Eqs.
\eqref{E:gamma_classical} and \eqref{E:Gamma_Cint_golden_rule} is
the same, and hence the desired result
$\Gamma_\text{cl}=\Gamma_\text{gr}$ is demonstrated: the quantum
spontaneous emission rate is exactly coincident with the decay rate
of the corresponding classical problem.

\section{Time evolution of the atomic operators} \label{sec:evol_atom_op}

To further characterize the dynamics of the two-level atom, in this
section we study the time evolution of the atomic operators and find
their vacuum expectation values using the Markov approximation. In
the Heisenberg picture, the time evolution of an operator $\op{A}$
that does not depend explicitly on time is
\begin{equation} \label{E:evolution_Heisenberg_picture}
 \frac{\mathrm{d} \op{A}}{\mathrm{d}t} = \frac{i}{\hbar} \left[\Ham,\op{A}\right].
\end{equation}
It is important to note here that even though the interaction
Hamiltonian depends explicitly on time the relevant commutation
relations are preserved for all $t$.

Using $\Ham = \Ham_\text{field}+\Ham_\text{at}+\Ham_\text{int}$ in
Eq. \eqref{E:evolution_Heisenberg_picture} it is easily found that
the time evolution of the atomic inversion operator is determined
by:
\begin{align}\label{E:evol_sigz}
 \frac{\mathrm{d} \op{\sigma}_z}{\mathrm{d}t} &=  \frac{2i}{\hbar} \op{\vec{F}}(\vec{r}_0'-\vec{v} t) \cdot (\tilde{\boldsymbol{\gamma}}_{eg}^\ast \op{\sigma}_+ - \tilde{\boldsymbol{\gamma}}_{eg} \op{\sigma}_-),
\end{align}
%
To make additional progress, in the next sub-section we characterize
the field operators.

\subsection{Evolution of the field operators}
\label{sec:evolution_sub}

Using again Eq. \eqref{E:evolution_Heisenberg_picture} and the
standard commutation relations, one finds that the annihilation
operators satisfy:
\begin{align}
 \frac{\mathrm{d} \cAnni_{n\k}}{\mathrm{d}t} &= - i\omega_{n\k} \cAnni_{n\k} + i \sqrt{\frac{ \omega_{n\k} }{2\hbar}} \hat{\vec{p}} \cdot \vec{F}_{n\k}^\ast(\vec{r}_0'-\vec{v} t
 ),
\end{align}
where we introduced the operator $ \hat{\vec{p}} =
(\tilde{\boldsymbol{\gamma}}_{eg}^\ast  \op{\sigma}_+ +
\tilde{\boldsymbol{\gamma}}_{eg} \op{\sigma}_-)$. This differential
equation can be written in an integral form as:
\begin{equation}
{{\hat a}_{n{\bf{k}}}}\left( t \right) = {{\hat a}_{n{\bf{k}}}}{e^{
- i{\omega _{n{\bf{k}}}}t}} + i\sqrt {\frac{{{\omega
_{n{\bf{k}}}}}}{{2\hbar }}} {\bf{F}}_{n{\bf{k}}}^*\left(
{{\vec{r}_0'} - {\bf{v}}t} \right) \cdot \int\limits_0^t {{\bf{\hat
p}}\left( {{t_1}} \right)u\left( {t - {t_1}} \right){e^{ - i
\omega_{n\k}' \left( {t - {t_1}} \right)}}} d{t_1}
\end{equation}
where $u$ denotes the Heaviside's step function. To make further
progress and obtain a closed form solution we use the Markov's
approximation \cite{allen_optical_1975}:
\begin{align}
\int\limits_0^t {{\bf{\hat p}}\left( {{t_1}} \right)u\left( {t -
{t_1}} \right){e^{ - i{\omega_{n\k}'}\left( {t - {t_1}} \right)}}}
d{t_1} & \approx  \tilde{\boldsymbol{\gamma}}_{eg} {\hat \sigma _ -
}\left( t \right)\int\limits_0^t {u\left( {t - {t_1}} \right){e^{ -
i\left( {\omega_{n\k}' - {\omega _0}}
\right)\left( {t - {t_1}} \right)}}} d{t_1} \nonumber \\
&+\tilde{\boldsymbol{\gamma}}_{eg}^\ast {\hat \sigma _ + }\left( t
\right)\int\limits_0^t {u\left( {t - {t_1}} \right){e^{ - i\left(
{\omega_{n\k}' + {\omega _0}} \right)\left( {t - {t_1}} \right)}}}
d{t_1}.
\end{align}
Moreover, for large $t$ one has the approximate identities
\begin{align}
\int\limits_0^t {u\left( {t - {t_1}} \right){e^{ - i\omega \left( {t
- {t_1}} \right)}}} d{t_1} \approx \int\limits_{ - \infty }^{ +
\infty } {u\left( {{t_1}} \right){e^{ - i\omega {t_1}}}} d{t_1}
\approx \pi \delta \left( \omega  \right),
\end{align}
where the imaginary part of the last integral is dropped. Hence,
from here we finally conclude that:
\begin{equation}\label{E:eq_evol_opAnni}
\cAnni_{n\k}(t) \approx \cAnni_{n\k}\e{-i \omega_{n\k} t} +  i\pi
\sqrt{\frac{\omega_{n\k} }{2\hbar}}
\vec{F}_{n\k}^\ast(\vec{r}_0')\e{i \k \cdot \vec{v} t }   \cdot
\left[ \tilde{\boldsymbol{\gamma}}_{eg}^\ast \op{\sigma}_+(t)
\delta\left( \omega_0 + \omega_{n\k}' \right) +
\tilde{\boldsymbol{\gamma}}_{eg} \op{\sigma}_-(t) \delta\left(
\omega_0 - \omega_{n\k}' \right)\right] .
 \end{equation}
Here it is worth pointing out that in Ref.
\cite{intravaia_quantum_2014} it was argued that the Markov
approximation might be inadequate to characterize the stationary
state regime ($t \to \infty$) in the presence of material
dissipation. In principle, our theory is not affected by such a
result (at least in the quasi-static limit considered later) because we deal with an ideal lossless system.

\subsection{The inversion operator expectation} \label{sec:evolution_Pe}
To determine the time evolution of the expectation of the inversion
operator, it is convenient to decompose the electromagnetic field
operator as $ \op{\vec{F}} = \op{\vec{F}}^- + \op{\vec{F}}^+$ such
that the annihilation part is
\begin{align} \label{E:Fminus}
\op{\vec{F}}^- (\vec{r},t)&= \sum_{\omega_{n\k}>0} \sqrt{\frac{\hbar
\omega_{n\k} }{2}}  \cAnni_{n\k}(t) \vec{F}_{n\k}(\vec{r}),
\end{align}
with $\cAnni_{n\k}(t)$ given by the approximate expression
\eqref{E:eq_evol_opAnni} and
$\op{\vec{F}}^+=\left(\op{\vec{F}}^-\right)^\dagger$. Calculating
now the expectation of both sides of Eq. \eqref{E:evol_sigz} and
using normal ordering
\begin{align}
\op{\vec{F}} \cdot (\tilde{\boldsymbol{\gamma}}_{eg}^\ast
\op{\sigma}_+ - \tilde{\boldsymbol{\gamma}}_{eg} \op{\sigma}_-) =
\op{\vec{F}}^+ \cdot (\tilde{\boldsymbol{\gamma}}_{eg}^\ast
\op{\sigma}_+ - \tilde{\boldsymbol{\gamma}}_{eg} \op{\sigma}_-)+
(\tilde{\boldsymbol{\gamma}}_{eg}^\ast \op{\sigma}_+ -
\tilde{\boldsymbol{\gamma}}_{eg} \op{\sigma}_-) \cdot \op{\vec{F}}^-
\end{align}
to eliminate the contribution of the free field part of the
electromagnetic field (first term in the right-hand side of Eq.
\eqref{E:eq_evol_opAnni}) \cite{allen_optical_1975}, it is found
after some simplifications that:
\begin{align}
 \frac{\mathrm{d} \left\langle  \op{\sigma}_z \right\rangle}{\mathrm{d}t} &=   2  \Gamma^- \left\langle\op{\sigma}_-\op{\sigma}_+\right\rangle    -2 \Gamma^+ \left\langle \op{\sigma}_+\op{\sigma}_-\right\rangle ,
\end{align}
where $\Gamma^+$ and $\Gamma^-$ are defined exactly in the same
manner as in Sec. \ref{sec:Fermi_golden_rule}. The above equation
assumes that the field is initially in its ground state. Using
relations $\op{\sigma}_z = 2\op{\sigma}_+\op{\sigma}_- - 1  $ and
$\op{\sigma}_-\op{\sigma}_+  = 1 - \op{\sigma}_+\op{\sigma}_- $, it is found that the time evolution of the excited state probability,
$\mathcal{P}_e \equiv \left\langle \op{\sigma}_+\op{\sigma}_-
\right\rangle$, is determined by
\begin{align}
  \frac{\mathrm{d} \mathcal{P}_e}{\mathrm{d}t}  &=      \Gamma^-  - \mathcal{P}_e \left( \Gamma^- +  \Gamma^+
  \right).
\end{align}
The general solution of this differential equation is
\begin{align} \label{E:Pe}
   \mathcal{P}_e(t)  &=  \frac{\Gamma^-}{\Gamma^-+\Gamma^+} \left( 1 - \e{-\left( \Gamma^-+\Gamma^+ \right) t } \right) + \mathcal{P}_e(0) \e{-\left( \Gamma^-+\Gamma^+ \right) t },
\end{align}
where $\mathcal{P}_e(0)$ is the initial excited state's probability.
In particular, this result shows that $\displaystyle - \left.
\frac{d\mathcal{P}_e/dt}{\mathcal{P}_e} \right|_{t=0}= \Gamma^+ -
\Gamma^-$ when $\mathcal{P}_e(0)=1/2$. Thus,
$\Gamma_{\rm{gr}}=\Gamma^+ - \Gamma^-$ determines the initial time
decay rate of the excited population in an ensemble of two-level
atoms with an identical number of atoms in the excited and ground
states at $t=0$. This shows that the total spontaneous emission rate
$\Gamma_{\rm{gr}}$ has a definite physical meaning also in the
quantum problem.

As expected, Eq. \eqref{E:Pe} predicts that in the absence of
relative motion (when $\Gamma^-=0$) the excited state probability
decays exponentially to zero, so that the atom decays to the ground
state $\ket{g}$ in a standard spontaneous emission  process and in
accordance with the results of section \ref{sec:Fermi_golden_rule}.
More interestingly, for velocities for which $\Gamma^-$ differs
significantly from zero, the probability of the excited state
evolves towards a stationary value $\mathcal{P}_{e,\infty} \equiv
\mathcal{P}_e(t=\infty)=\Gamma^-/(\Gamma^-+\Gamma^+)$ that depends
on the relative strength of the decay rates $\Gamma^+$, $\Gamma^-$
and that determines some sort of dynamic equilibrium of the
two-level atom. Note that in the stationary regime ($t\to\infty$) it
is possible to write $\mathcal{P}_{e,\infty}{\Gamma ^ + } - \left(
{1 - \mathcal{P}_{e,\infty}} \right){\Gamma ^ - } = 0$, such that
the number of transitions to the excited state induced by the
relative motion $\left( {1 - \mathcal{P}_{e,\infty}} \right){\Gamma
^ -}$ equals the number of transitions to the ground state induced
by the emission processes $\mathcal{P}_{e,\infty}{\Gamma ^ + }$.

The picture that emerges from this result is that the limit
$t\to\infty$ may have a dynamical character, such that energy is
continuously extracted from the kinetic degrees of freedom and
emitted as light. Indeed, the fact that $\mathcal{P}_{e,\infty}$ is
nonzero proves that in the Schr\"{o}dinger picture the atomic wave
function is a mixed state for large $t$, and the expectation of the
dipole moment is nonzero for any mixed state ($\left\langle
\op{\sigma}_- \right\rangle \ne 0$).

Notably, when the classical decay rate
$\Gamma_\text{cl}=\Gamma_\text{gr}=\Gamma^+ - \Gamma^-$ is negative,
i.e., the transitions with negative Doppler shifted frequency
dominate over the emission processes with positive Doppler shifted
frequency, then $\mathcal{P}_{e,\infty}> 1/2$. In this situation, a
population inversion takes place, independently of the initial
atomic state.

Importantly, a two-level atom is a system with an intrinsic
saturation mechanism, because its energy is bounded from above.
Moreover, the dipole moment amplitude is also bounded, and reaches
the maximum for mixed-states with $\mathcal{P}_{e,\infty} = 1/2$,
i.e., at the onset of the classical instability. Hence, different
from the classical case, in the quantum problem the dipole moment
cannot grow exponentially, and the maximum oscillation amplitude is
reached for $\mathcal{P}_{e,\infty} = 1/2$. This suggests that
classical situations leading to strong instabilities should
correspond in the quantum case to $\mathcal{P}_{e,\infty}$
marginally larger than $1/2$. In section \ref{sec:Numerical_results}
it will be shown with a numerical example that this is indeed the
case. The fact that $\mathcal{P}_{e,\infty}
> 1/2$ suggests that if a ladder of energy levels would be accessible
then the dipole oscillations would grow exponentially in time,
similar to the classical problem.

In summary, the picture that emerges from our quantum optics
analysis is that the two-level atom evolves towards a dynamical
equilibrium state, such that the number of transitions in the
ascending direction equals the number of transitions in the
descending direction.

\subsection{Friction force} \label{sec:friction_force}

As discussed in section \ref{sec:Fermi_golden_rule}, the transitions
associated with oscillators with negative frequencies must be
accompanied by a conversion of kinetic energy into light. The
optical friction force responsible for this exchange of energy can
be calculated in the frame co-moving with the two-level atom using
${\mathcal{F}}_x = {\p}_e \cdot
\partial_{x'} \E' = {\p} \cdot
\partial_{x'} \bf{F}'$, where the partial derivative acts
over the spatial coordinate along the direction of motion
\cite{intravaia_quantum_2014}. Promoting all the relevant physical
quantities to operators and using normal ordering, it is follows
that the quantum expectation of the force is:
\begin{align}
\left\langle \mathcal{F}_x (\vec{r}_0')\right\rangle &= 2 \,
\mathrm{Re} \left\{ \left\langle \op{\p} \cdot \partial_{x'} \vec{
\hat F}'^-(\vec{r}_0') \right\rangle \right\}.
\end{align}
Using Eqs. \eqref{E:eq_evol_opAnni}-\eqref{E:Fminus} and assuming
that at initial time the field is in the vacuum state  it follows
that in the non-relativistic limit the force is given by:
\begin{align}
\left\langle \op{\mathcal{F}}_x(\vec{r}_0') \right\rangle =  &
\sum_{\omega_{n\k}>0}   \mathrm{Re}\left\{ i\pi \omega_{n\k}
\tilde{\boldsymbol{\gamma}}_{eg}^\ast  \cdot \left[
\partial_{x'}\vec{F}_{n\k}(\vec{r}_0') \otimes
\vec{F}_{n\k}^\ast(\vec{r}_0')  \right] \cdot
\tilde{\boldsymbol{\gamma}}_{eg}  \delta\left( \omega_0 -
\omega_{n\k}' \right)  \left\langle \op{\sigma}_+
\op{\sigma}_-\right\rangle \right\} \nonumber \\ &  +
\sum_{\omega_{n\k}>0}  \mathrm{Re}\left\{ i\pi \omega_{n\k}
\tilde{\boldsymbol{\gamma}}_{eg}  \cdot \left[
\partial_{x'}\vec{F}_{n\k}(\vec{r}_0') \otimes
\vec{F}_{n\k}^\ast(\vec{r}_0')  \right] \cdot
\tilde{\boldsymbol{\gamma}}_{eg}^\ast  \delta\left( \omega_0 +
\omega_{n\k}' \right)  \left\langle \op{\sigma}_-
\op{\sigma}_+\right\rangle \right\}.
\end{align}
The system is invariant to translations along the direction of
motion and thus for the mode $n\k$ one has $\partial_{x'} = i k_x$,
being $k_x$ the $x$-component of the wave vector. Hence, writing the
expectation of the atomic operators in terms of $\mathcal{P}_e$, one
finally finds that the friction force is
\begin{align} \label{E:friction_force}
\left\langle \op{\mathcal{F}}_x(\vec{r}_0') \right\rangle =  & -\mathcal{P}_e \sum_{\omega_{n\k}>0}   \mathrm{Re}\left\{ \pi   k_x \omega_{n\k}  \tilde{\boldsymbol{\gamma}}_{eg}^\ast  \cdot \left[  \vec{F}_{n\k}(\vec{r}_0') \otimes \vec{F}_{n\k}^\ast(\vec{r}_0')  \right] \cdot  \tilde{\boldsymbol{\gamma}}_{eg}  \delta\left( \omega_0 - \omega_{n\k}' \right)  \right\}  \nonumber  \\
& -\left(1 - \mathcal{P}_e  \right) \sum_{\omega_{n\k}>0}     \mathrm{Re}\left\{ \pi  k_x \omega_{n\k}        \tilde{\boldsymbol{\gamma}}_{eg}  \cdot \left[ \vec{F}_{n\k}(\vec{r}_0') \otimes \vec{F}_{n\k}^\ast(\vec{r}_0')  \right] \cdot  \tilde{\boldsymbol{\gamma}}_{eg}^\ast  \delta\left( \omega_0 + \omega_{n\k}' \right)   \right\}.
\end{align}
It is demonstrated in the next section that in the quasi-static
limit the friction force can be directly written in terms of the
transition rates $\Gamma^-$ and $\Gamma^+$.


\section{The quasi-static limit} \label{sec:Numerical_results}

\subsection{Analytical development}

To illustrate the concepts developed throughout this paper and link
them to the classical results of Ref.
\cite{silveirinha_optical_2014}, next we derive explicit formulas
for the spontaneous emission rates and friction force using a
quasi-static approximation for the quantized electromagnetic fields.

In a quasi-static approximation the fields are purely electric
$\vec{F}_{n\k} \approx  \left( \vec{E}_{\k} \quad \vec{0} \right)^T
$ \cite{maier_plasmonics:_2010}, and the electric field
$\vec{E}_{\k} = -\nabla \phi_{\k}$ is written in terms of an
electric potential that satisfies $\nabla \cdot \left( {\varepsilon
\left( {\omega ,z} \right)\nabla {\phi _{{\bf{k}}}}} \right) = 0$.
Assuming that the plasmonic slab is located in $z<0$ region (so that
the metal-air boundary is at $z=0$), it follows that the electric
potential must be of the form
\begin{align}
{\phi _{\bf{k}}} = {A_{{{\bf{k}}}}}{e^{i{{\bf{k}}_{||}} \cdot
{\bf{r}}}}{e^{ - {k_{||}}\left| z \right|}}, \end{align}
where $\k_\parallel= k_x \hat{\vec{x}} + k_y \hat{\vec{y}}$ is the
wave vector (here, we use the subscript $\parallel$ to highlight
that the wave vector is parallel to the interface) and $A_{\bf{k}}$
is a normalization constant. Moreover, the relevant eigenfrequency
is ${\omega _{\bf{k}}} = \omega_\text{sp}$  where $\omega_\text{sp}$
corresponds to the surface plasmon resonance $\varepsilon \left(
{{\omega _{sp}}} \right) = - {\varepsilon _0}$ .

To determine the value of $A_{\k}$, we use the normalization
condition \eqref{E:relation_normalization_modes} that gives
\begin{align} \label{E:normalization_electrostatic_field}
 |A_{\k}|^2 \int d^3\vec{r}~    {k_\parallel^2} \e{-2  k_\parallel |z| } \frac{\partial~ \left[\omega \eps(\omega,z)\right]}{\partial \omega}  = 1.
\end{align}
Considering that the permittivity of the plasmonic slab is modeled
by a lossless Drude model, $\varepsilon  = {\varepsilon _0}\left( {1
- 2\omega _{{\rm{sp}}}^2/{\omega ^2}} \right)$, one has
\begin{equation}
\frac{\partial \,  \left[\omega \eps(\omega,z)\right]  }{\partial
\omega} \xrightarrow[]{\omega = \omega_\text{sp}} \left\{
\begin{tabular}{c r}
$\eps_0,$  &\quad $z>0$ \\
 $3\eps_0,$ &\quad $z<0$
\end{tabular}
\right.
\end{equation}
and hence \eqref{E:normalization_electrostatic_field} gives that
\begin{align}
|A_{\k}|=\sqrt{\frac{1}{2 k_\parallel \eps_0 S_0  }},
\end{align}
where $S_0$ is surface area of the plasmonic slab.

Substituting the quasi-static electric field into Eqs.
\eqref{E:gammaPlus}, \eqref{E:gammaMoins},
\eqref{E:Im_Cint_eigenmodes} and \eqref{E:friction_force}, it is now straightforward to determine
the spontaneous emission rates, the interaction's dyadic and the friction force. For
simplicity, in what follows it is supposed that the electric dipole
is oriented along the $z$-direction so that
$\boldsymbol{\gamma}_{eg} = \gamma_{eg} \hat{\vec{z}}$. In this case
the transition rate to the ground state \eqref{E:gammaPlus}
simplifies to:

\begin{equation}
 \Gamma^+ =    \sum_{\k_{||}}   \frac{{\pi {\omega _{{\rm{sp}}}}}}{\hbar }\frac{{{k_{||}}}}{{2{\varepsilon _0}{S_0}}}{\left| {\gamma _{eg}^{}} \right|^2}  \e{- 2 k_\parallel d }  \delta\left( \omega_\text{sp}+ k_x  v -\omega_0
 \right),
\end{equation}
where $d$ is the distance between the two-level atom and the
plasmonic slab. The sum over $\k_\parallel$ can be transformed into
an integral by substituting $ \frac{1}{S_0 }\sum_{\k_\parallel} \to
\frac{1}{(2\pi)^2} \iint dk_x dk_y~  $. This gives:
\begin{align} \label{E:gamma_p_quasi_static}
 \Gamma^+ &=    \frac{2 |\gamma_{eg}|^2}{ \eps_0 \hbar } \frac{1}{d^3} \,  G\left( \frac{\omega_0 d}{|v|}, \frac{\omega_\text{sp} d}{|v|} \right),
\end{align}
where we introduced the function
\begin{align}
G\left(a,b \right) =\frac{ b  }{8\pi} \int_0^{\infty}    \sqrt{u^2 +
\left( a - b \right)^2} \e{-2 \sqrt{u^2 + \left( a - b \right)^2} }
~\mathrm{d}u.
\end{align}
Similarly, using Eq. \eqref{E:gammaMoins} it is readily verified that
the rate of transitions to the excited state is
\begin{align} \label{E:gamma_m_quasi_static}
 \Gamma^- &=    \frac{2 |\gamma_{eg}|^2}{ \eps_0 \hbar } \frac{1}{d^3} \,  G\left( \frac{-\omega_0 d}{|v|}, \frac{\omega_\text{sp} d}{|v|}
 \right).
\end{align}
Now, using the fact that $\Gamma_\text{cl}=\Gamma^+ - \Gamma^-$
and Eq. \eqref{E:gamma_classical} one sees that the $zz$ component
of the interaction dyadic satisfies:
\begin{align}
{\mathop{\rm Im}\nolimits} \left\{ {{C_{{\mathop{\rm int}} ,zz}}}
\right\} = \frac{1}{{{d^3}}}\left[G\left( \frac{\omega_0 d}{|v|},
\frac{\omega_\text{sp} d}{|v|}
 \right) -G\left( \frac{-\omega_0 d}{|v|},
\frac{\omega_\text{sp} d}{|v|}
 \right) \right]
 \end{align}
This result is exactly coincident with Eq. (15) of Ref.
\cite{silveirinha_optical_2014}, which was derived using a totally
different approach. This agreement provides an independent check of
the theoretical concepts introduced in this article. It should be
noted that in the quasi-static approximation, the free space's
contribution to the interaction constant is lost
($\Im{\db{C}_\text{tot}} = \Im {\db{C}_\text{int} }$). However, as
demonstrated in \cite{silveirinha_optical_2014}, this contribution
is tiny as compared to the imaginary part of ${{C_{{\mathop{\rm
int}} ,zz}}}$ and hence can be safely neglected.

On the other hand, it can be shown that within the same
approximations the friction force acting on the two-level atom
\eqref{E:friction_force} simplifies to
\begin{align}
\left\langle \op{\mathcal{F}}_x  \right\rangle & =  -\mathcal{P}_e
\Gamma^+ \hbar \frac{\omega_0 - \omega_\text{sp}}{v}    +
\left(1-\mathcal{P}_e \right) \Gamma^-   \hbar   \frac{\omega_0 +
\omega_\text{sp}}{v},
\end{align}
and in the stationary state ($t \to \infty$) it further reduces  to
\begin{align}
\left\langle \op{\mathcal{F}}_{x,\infty}  \right\rangle & =
\frac{2\hbar\omega_\text{sp}}{v}  \frac{\Gamma^+  \Gamma^-
}{\Gamma^+ + \Gamma^-}.
\end{align}
Note that the signs of the friction force and atom velocity are
opposite because the velocity of the atom is $-v$ in the reference
frame of the metal slab ($v$ represents the relative velocity of the
metal with respect to the atom). Interestingly, the above expression
shows that whenever the rate of ascending transitions $\Gamma^-$ is
nonzero then the friction force will also be nonzero, even if the
corresponding classical problem is stable (i.e.
$\Gamma^+>\Gamma^-$). This situation contrasts with the case of
sliding \emph{lossless} slabs, where a friction force implies an
unstable response \cite{silveirinha_theory_2014}. The difference is
that in the latter problem the ``emitter'' is infinitely extended in
space and hence without absorption the emitted energy will
necessarily build up, leading to an exponential growth. On the other
hand, for a moving atom the emitted light can be radiated away and
thus it does not necessarily increase the stored energy in the
vicinity of the atom.

Similarly, the energy emitted per unit of time is given by
$\left\langle \op{\mathcal{F}}_{x,\infty} \right\rangle v = 2\hbar
\omega_\text{sp} \mathcal{P}_{e,\infty}{\Gamma ^ + }$ (in agreement
with the transition rates evaluated in section
\ref{sec:evolution_Pe}; the leading factor of ``2'' can be
understood noting that both the ascending and descending transitions
emit a plasmon). Hence, the emitted power saturates at some value
rather than growing exponentially as in the classical case.

\subsection{Numerical study}

\begin{figure*}[!ht]
\centering
\includegraphics[width=.9\linewidth]{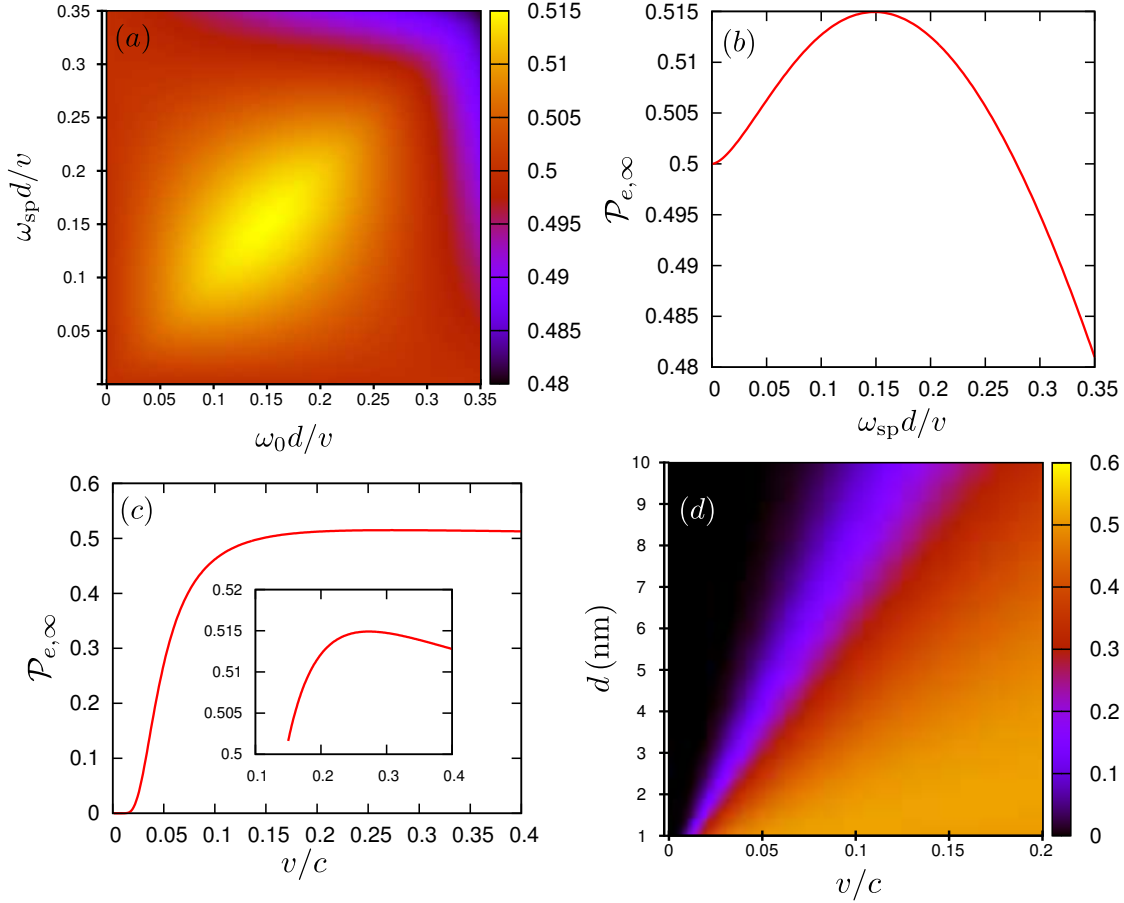}
         \caption{Plot of  the probability $\mathcal{P}_{e,\infty}$ in different situations: (a)  Density plot of $\mathcal{P}_{e,\infty}$ as a function of the dimensionless
         parameters $\omega_0 d/v$ and $\omega_\text{sp} d/v$. (b) $\mathcal{P}_{e,\infty}$ as a function of $\omega_\text{sp} d/v$ for $\omega_0 = \omega_\text{sp}$.
         (c) $\mathcal{P}_{e,\infty}$ as a function of the velocity $v$ for a two-level atom located at a distance $d=3~ \nano\meter$ from a lossless silver slab
          with $\omega_\text{sp}/(2 \pi)= 646 ~\tera\hertz$ and an atomic transition frequency $\omega_0=\omega_\text{sp}$.
          The inset shows a zoom of the same curve near $\mathcal{P}_{e,\infty}=1/2$.
         (d) Density plot of $\mathcal{P}_{e,\infty}$ as a function of the velocity and of the distance, for $\omega_\text{sp}/(2 \pi)= 646 ~\tera\hertz$ and $\omega_0=\omega_\text{sp}$.}
\label{fig:densityplot_pPlus}
\end{figure*}

We are now ready to study how the dynamic equilibrium state
$\mathcal{P}_{e,\infty}=\Gamma^-/(\Gamma^-+\Gamma^+)$ induced by the
interaction with the SPPs varies with the relative velocity. The
excited state probability in the limit $t \to \infty$ depends only
on the normalized parameters $\omega_0 d/|v|$ and $\omega_\text{sp}
d/|v|$. A density plot of $\mathcal{P}_{e,\infty}$ as a function of
the two normalized variables is represented in Fig.
\ref{fig:densityplot_pPlus} (a) in a parametric range near to the
maximum of the function.
As seen, the maximum value of $\mathcal{P}_{e,\infty}$ occurs
exactly for $\omega_0=\omega_\text{sp}$, i.e., when the atomic
transition frequency is coincident with the SPP resonance.
Furthermore, as expected, for small velocities (which correspond to
large values of the normalized parameters) the value of
$\mathcal{P}_{e,\infty}$ is negligible. A detailed variation of
$\mathcal{P}_{e,\infty}$ in the interesting case
$\omega_0=\omega_\text{sp}$ is depicted in Fig.
\ref{fig:densityplot_pPlus} (b). The optimal value that maximizes
$\mathcal{P}_{e,\infty}$ is $\omega_\text{sp}d/v=0.148$ and
corresponds to the optimal velocity $v \approx 6.72~
\omega_\text{sp}d $.

Notably, there is a range of parameters for which
$\mathcal{P}_{e,\infty}>1/2$ implying a negative spontaneous
emission $\Gamma_{\rm{gr}}<0$. Moreover, consistent with the
discussion of Sec. \ref{sec:evolution_sub}, in such a regime the
maximum value (about $0.515$) reached by $\mathcal{P}_{e,\infty}$ is
only marginally larger than $1/2$. This ensures that the dipole
oscillation amplitude in the dynamical equilibrium has the largest
possible value, when the associated classical problem is
characterized by an unstable response. Indeed, whereas in the
classical case \cite{silveirinha_optical_2014} the dipole moment
amplitude may reach arbitrarily large values (in the linear regime),
in a two-level atom it saturates for mixed states with
$\mathcal{P}_{e} \approx 1/2$.

It is interesting to look at an example with a realistic material
with more detail. Here, as in Ref. \cite{silveirinha_optical_2014}
we consider that the metal is silver, which is assumed to be modeled
by a Drude model with a plasma frequency $\omega_\text{sp}/(2 \pi)=
646 ~\tera\hertz$ \cite{johnson_optical_1972}. The effect of
metallic loss is not included in our calculation, but as shown in
Ref. \cite{silveirinha_optical_2014} it does not change the
qualitative picture, and only acts to reduce the strength of the
negative spontaneous emission rate.

Figure \ref{fig:densityplot_pPlus} (c) shows the variation of
$\mathcal{P}_{e,\infty}$ with the relative velocity for a distance
$d=3~ \nano\meter$ between the two-level atom and the silver slab,
and an atomic transition frequency $\omega_0=\omega_\text{sp}$. As
seen for low velocities (typically $v<0.05 c$),
$\mathcal{P}_{e,\infty}$ remains near zero meaning that the
processes involving negative Doppler shifted frequency are
negligible in agreement with the conclusions of Secs.
\ref{sec:Fermi_golden_rule} and \ref{sec:evol_atom_op}. Crucially,
for higher velocities the situation changes, and as we enter in the
range of parameters shown in Fig. \ref{fig:densityplot_pPlus}
(a)-(b), the excited state probability in the dynamical equilibrium
reaches a value around 50 \%. In addition, as can be seen in the
inset, the probability reaches a maximum for the velocity $v \approx
0.273 c $.

To have a clear idea of the range of parameters needed for a
negative spontaneous emission we represent in Fig.
\ref{fig:densityplot_pPlus} (d) a density plot of
$\mathcal{P}_{e,\infty}$ as a function of the velocity and of the
distance, for the same scenario as in the previous example.
Obviously, increasing the distance between the atom and the slab
weakens the interaction, and in particular the effect of the quantum
oscillators with negative Doppler shifted frequencies. As a
consequence, the threshold velocity to obtain a population inversion
is higher. We note that the effects of time-retardation and
relativistic corrections are not expected to change the general
conclusions of the article, as it was shown in Ref.
\cite{silveirinha_optical_2014} that they generally correspond to
small corrections in the classical case.

\begin{figure*}[!ht]
\centering
\includegraphics[width=.9\linewidth]{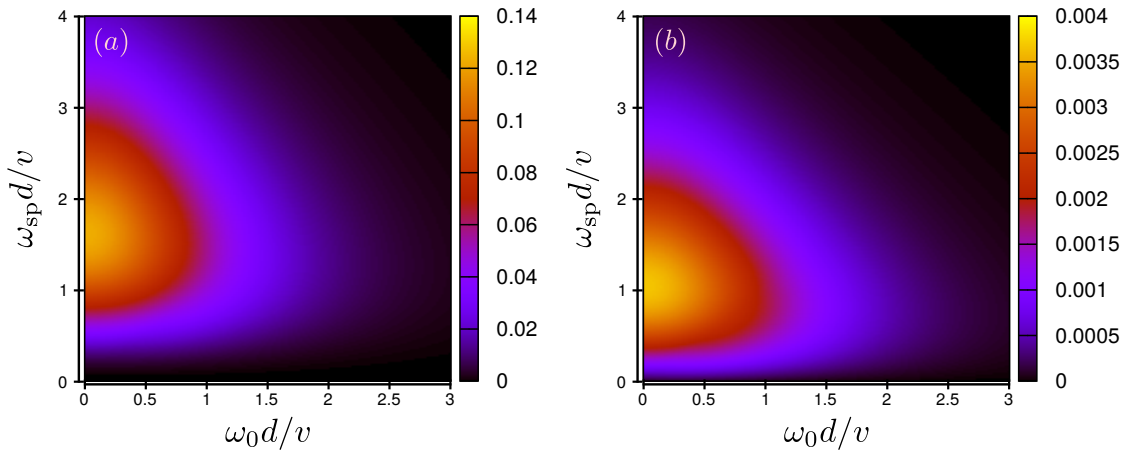}
         \caption{Plot of the expectation value of (a) the normalized friction force $  \left\langle  \mathcal{F}_{x,\infty} \right\rangle \cdot \frac{4 \pi \eps_0 d^4}{|\boldsymbol{\gamma}_{eg}|^2}  $  and (b) the normalized radiated power $ \left\langle  \mathcal{P} \right\rangle \cdot  \frac{4 \pi \eps_0 d^3}{\omega_\text{sp}|\boldsymbol{\gamma}_{eg}|^2}  $ as a function of the dimensionless parameters  $\omega_0 d/v$ and $\omega_\text{sp} d/v$.}
\label{fig:friction_power}
\end{figure*}

Next, we characterize the quantum friction force. Figure
\ref{fig:friction_power} shows the equilibrium friction force
$\left\langle \op{\mathcal{F}}_{x,\infty} \right\rangle$ as well as
the power radiated by the system $\left\langle
\op{\mathcal{F}}_{x,\infty}  \right\rangle v$ as a function of the
normalized parameters $\omega_0 d/|v|$ and $\omega_\text{sp}d/|v|$.
As expected, the force and the emitted power are significant only
for sufficiently high velocities and small distances. Comparing the
plot of $\mathcal{P}_{e,\infty}$ (Fig. \ref{fig:densityplot_pPlus}
(a)) with the plots in Fig. \ref{fig:friction_power}, it is somewhat
surprising that the maxima of $\mathcal{P}_{e,\infty}$ and
$\left\langle \mathcal{F}_{x,\infty} \right\rangle$ occur in rather
different regions of the parameter space. Indeed, the maxima of the
friction force and power occur for $\omega_0 \ll \omega_{\rm{sp}}$,
whereas the maximum of $\mathcal{P}_{e,\infty}$ occurs for $\omega_0
= \omega_{\rm{sp}}$. Yet, it is possible to see from equations
\eqref{E:gamma_p_quasi_static} and \eqref{E:gamma_m_quasi_static}
that a case with $\omega_0 \ll \omega_{\rm{sp}}$ corresponds to
$\Gamma^+ \approx \Gamma^-$ and hence to $\mathcal{P}_{e,\infty}
\approx 1/2$. Thus, even though the maximum of friction force does
not fall within the region wherein $\mathcal{P}_{e,\infty}$ exceeds
$1/2$, the value of $\mathcal{P}_{e,\infty}$ is actually only
marginally smaller than the maximum. For example, at the maximum of
the friction force ($w_0 d/v = 0^-$ and $w_{\rm{sp}} d/v \approx 1.62$) one has $\mathcal{P}_{e,\infty}=1/2+0^-$. The key
point that shows there is no lack of consistency is that if the atom
had an infinite number of energy levels the quantum friction would
likely grow exponentially with time, and hence in such a case the
peak friction force would necessarily fall within the region wherein
the total spontaneous emission rate is negative.

We verified that within our model the friction force becomes
exponentially small in the $v \to 0$ limit. This property is
consistent with a result obtained by Pendry for lossless plasmonic
slabs in a shear motion \cite{pendry_quantum_2010}. In contrast, for
dissipative materials with a nonzero conductivity in the static
limit, the friction force exhibits a power law dependence on the
velocity
\cite{intravaia_quantum_2014,Henkel_2015,hoye_casimir_friction_2015}.
Indeed, it is important to highlight that in our problem the
friction force is exclusively due to the conversion of kinetic
energy into electromagnetic radiation (plasmons) because the system
is lossless. This process is particularly efficient when
$\omega_{\rm{sp}} d/v \sim 1$ (see Fig. \ref{fig:friction_power}).
On the other hand, lossy materials provide additional channels for
dissipation, namely enable the conversion of kinetic energy into
heat, and this explains why the friction force is stronger in the $v
\to 0$ limit in dissipative systems.



\section{Conclusion}

It was demonstrated that the rate of spontaneous emission by a
moving two-level atom interacting with the near-field of a plasmonic
slab is, in the most general case, determined by two concurrent
processes: the conservative ones involving only positive Doppler
shifted frequencies and the non-conservatives involving only
negative Doppler shifted frequencies. The transitions associated
with negative Doppler shifted frequencies are due to the conversion
of kinetic energy into light \cite{meyer_quantum_1985}.

It was proven that when the non-conservative processes are dominant,
the system is characterized by a negative spontaneous emission
meaning that the transition rate to the excited state exceeds the
transition rate to the ground state. In this regime, the
corresponding classical problem is characterized by an unstable
response. Different from the classical situation, in the quantum
case the dipole moment oscillations are intrinsically bounded. It
was shown with a numerical example that the unstable regime
corresponds in the quantum system to a saturation of the dipole
moment amplitude.

The links between the quantum theory presented here and the
classical theory of Ref. \cite{silveirinha_optical_2014} have been
emphasized. In particular, it was shown that the classical decay
rate is exactly coincident with the quantum spontaneous emission
rate. Using a quasi-static approximation for the electromagnetic
field, the typical range of distances and velocities to observe
negative spontaneous emission and an inversion of population have
been detailed. Additionally, we studied the phenomenon of quantum
friction. Different from its classical counterpart
\cite{silveirinha_optical_2014}, the quantum friction effect in the
considered system has no velocity threshold and in particular may
occur even when the total spontaneous emission rate remains
positive, i.e., when the classical system is stable. Moreover, the
regime wherein the friction force is maximal does not overlap with
the regime wherein the spontaneous emission rate is negative. Thus,
even though the stability threshold of the classical problem is
coincident with the threshold of negative spontaneous emission, the
physics of the classical and quantum systems is generally rather
different mainly due to the unique energy spectrum of the two-level
atom which prevents the development of instabilities in the quantum
case.

\begin{acknowledgments}
This work was partially funded by Funda\c{c}\~{a}o para Ci\^{e}ncia
e a Tecnologia under project PTDC/EEI-TEL/4543/2014.
\end{acknowledgments}

\appendix

\section{The field radiated by a moving electric dipole} \label{sec:appendix_field_dipole}

In the following, we calculate the electromagnetic field radiated by
a moving electric dipole in the reference frame co-moving with the
plasmonic slab. The electromagnetic field satisfies the Maxwell's
equations
\begin{align} \label{E:Maxwell_equation_time}
  \op{N} \vec{F} = i \frac{\partial}{\partial t}  \vec{M} \cdot \vec{F} + i\vec{J},
\end{align}
where  $\vec{J}$ is a six-vector representing the current density
due to the moving source. For a moving dipole it is of the form:
\begin{align}
  \vec{J}(\vec{r},t)= \frac{d }{d t} \left[ \delta(\vec{r}-\vec{r}_0(t)) \p(t) \right],
\end{align}
where $\vec{r}_0(t)=\vec{r}_0' - \vec{v} t$ are the coordinates of
the two-level atom in the reference frame of the plasmonic slab, and
$\p=\left( \p_e \quad \p_m \right)^T$ is the generalized dipole
moment, such that $\p_e$ and $\p_m$ represent the electric and
magnetic dipole moments (the magnetic dipole moment is included here
only for the sake of generality). The radiated field can be written
in terms of the temporal Green-function as
\begin{align} \label{E:field_scattered_Green}
\vec{F}(\vec{r},t) =&  \iint \db{G}(\vec{r},\vec{r}_1,t-t_1) \cdot
\vec{J}(\vec{r}_1,t_1) d\vec{r}_1dt_1 ,
\end{align}
where $\db{G}(\vec{r},\vec{r}_1,t)$ is the solution of
\begin{align} \label{E:Maxwell_Green's_function}
 \left( \op{N} - i  \frac{\partial }{\partial t} \vec{M} \right) \cdot \db{G}(\vec{r},\vec{r}_1,t) = i \delta(\vec{r}-\vec{r}_1)\delta(t) {\bf{1}}_{6\times6}
\end{align}
that satisfies the causality condition
$\db{G}(\vec{r},\vec{r}_1,t)=0$ for $t<0$. Assuming that the time
dependence of the dipole moment is of the form $\p(t)=\p {e^{ -
i\omega t}}$ it is readily found that:
\begin{equation} \label{E:F_dip}
{\bf{F}}\left( {{\bf{r}},t} \right) = \frac{d}{{dt}}\left[ \int
\db{G} {\left( {{\bf{r}},{{\bf{r}}_0}\left( t_1 \right),t - {t_1}}
\right) \cdot {\bf{p}}{e^{ - i\omega {t_1}}} d{t_1}} \right].
\end{equation}

In Appendix \ref{sec:appendix_Green's_function}, it is shown that the
Green's function in the spectral domain can be expanded in terms of
the electromagnetic eigenmodes $\vec{F}_{n\k}$ as in Eq.
\eqref{E:green_function_spectral}, so that in the time domain:
\begin{equation} \label{E:green_function_time}
 \db{G}(\vec{r},\vec{r}_1, t)= -u\left( t \right) \frac{1}{2}\sum_{n\k} {e^{ - i{\omega _{n{\bf{k}}}}t}} \vec{F}_{n\k}(\vec{r}) \otimes
 \vec{F}_{n\k}^\ast(\vec{r}_1),
\end{equation}
where $\otimes$ denotes the tensor product of two six-vectors, and
$u\left( t \right)$ is the Heaviside step function. The eigenmodes
are normalized as in Eq. \eqref{E:relation_normalization_modes}, and
the sum includes modes with positive, negative and zero frequencies.
Substituting this result into Eq. \eqref{E:F_dip} one obtains:
\begin{align}
\vec{F}(\vec{r},t) =&     \sum_{n\k}  \frac{ \omega  -\k\cdot
\vec{v}  }{2\left( \omega_{n\k}' - \omega   \right) }
\vec{F}_{n\k}(\vec{r}+ \vec{v} t) \otimes
\vec{F}_{n\k}^\ast(\vec{r}_0') \cdot \vec{p}~ \e{-i \omega   t},
\end{align}
where $\omega_{n\k}'=\omega_{n\k}  + \k \cdot \vec{v}$ is the
Doppler shifted frequency. Using the following completeness relation
(which can be derived using ideas analogous to those described in
Appendix \ref{sec:appendix_Green's_function}; $\vec{M}_\infty \equiv
{\lim _{\omega  \to \infty }}  \vec{M} ({\bf{r}},\omega)$)
\begin{equation}
\sum_{n\k} \frac{1}{2}  \vec{F}_{n\k}(\vec{r}) \otimes
\vec{F}_{n\k}^\ast(\vec{r}_1) = \vec{M}^{-1}_{\infty}
\delta(\vec{r}-\vec{r}_1),
\end{equation}
and supposing that the dipole is in the free-space region (with
material matrix $\vec{M}_0$), it is found that:
\begin{equation}
{\bf{F}} = {e^{ - i\omega t}}\sum\limits_{n{\bf{k}}}
{\frac{1}{2}\frac{{{\omega _{n{\bf{k}}}}}}{{\left( { \omega_{n\k}' -
\omega } \right)}}{{\bf{F}}_{n{\bf{k}}}}\left( {{\bf{r}} +
{\bf{v}}t} \right)} {\bf{F}}_{n{\bf{k}}}^*\left( \vec{r}_0' \right)
\cdot {\bf{p}} - {e^{ - i\omega t}}{\bf{M}}_0^{ - 1} \cdot
{\bf{p}}\delta \left( {{\bf{r}} + {\bf{v}}t - \vec{r}_0'} \right).
\end{equation}
The reality of the electromagnetic field implies that the above
summation can be restricted to eigenmodes with positive frequencies
as follows:
\begin{align} \label{E:Frad_final}
\vec{F}(\vec{r},t) = &  \sum_{\omega_{n\k}>0} \frac{\omega_{n\k}}{2
} \left( \frac{1 }{ \omega_{n\k}' -\omega}
\vec{F}_{n\k}(\vec{r}+\vec{v}t) \otimes
\vec{F}^\ast_{n\k}(\vec{r}_0') + \frac{1 }{ \omega_{n\k}' +\omega }
\vec{F}^\ast_{n\k}(\vec{r}+\vec{v}t) \otimes
\vec{F}_{n\k}(\vec{r}_0')  \right) \cdot \p \e{-i\omega t}  \nonumber  \\
 & - \delta(\vec{r}+\vec{v}t-\vec{r}_0') \vec{M}_0^{-1}
\cdot \p \e{-i\omega t}.
\end{align}
Equation \eqref{E:Frad_final} gives the exact solution for the
fields radiated by the moving dipole in the reference frame of the
plasmonic slab. Similar to Sec. \ref{sec:formalism}, to obtain the
corresponding fields in the frame co-moving with the dipole we use
simply the non-relativistic approximation
${\vec{F}}'(\vec{r}',t')\approx {\vec{F}}(\vec{r}'-\vec{v} t,t )$.
Note that  $\p'(t')\approx \p(t)$ within the same non-relativistic
approximation and hence the frequency of oscillation of the dipole
does not need to be transformed. Therefore, one concludes that
\begin{align} \label{E:F_dip_final}
\vec{F}'(\vec{r}',t) \approx &   \sum_{\omega_{n\k}>0}
\frac{\omega_{n\k}}{2 } \left( \frac{1 }{ \omega_{n\k}' -\omega}
\vec{F}_{n\k}(\vec{r}') \otimes  \vec{F}^\ast_{n\k}(\vec{r}_0') +
\frac{1 }{ \omega_{n\k}' +\omega }  \vec{F}^\ast_{n\k}(\vec{r}')
\otimes  \vec{F}_{n\k}(\vec{r}_0')  \right)    \cdot \p \e{-i\omega
t}  \nonumber \\
& -   \delta(\vec{r}'-\vec{r}_0') \vec{M}_0^{-1} \cdot \p
\e{-i\omega t}.
\end{align}
%

\section{Eigenmode expansion of the Green's function in the spectral domain} \label{sec:appendix_Green's_function}

In this appendix, we derive an explicit formula for the Green's
function in the frequency domain  $\db{G} =
\db{G}(\vec{r},\vec{r}_1, \omega)$, which from Eq.
\eqref{E:Maxwell_Green's_function} satisfies:
\begin{align} \label{E:Maxwell_Green's_function_freq}
 \left( \op{N} - \omega \vec{M} \right) \cdot \db{G}(\vec{r},\vec{r}_1, \omega) = i \delta(\vec{r}-\vec{r}_1)
 {\bf{1}}_{6\times6}.
\end{align}
It is well known that the Green's function can be expanded as a sum
of eigenvectors when all the involved materials are dispersionless
\cite{novotny_principles_2012,sakoda_optical_2004}. In contrast, in
presence of material dispersion the application of the spectral
theorem is not direct because the relevant differential operators
are not Hermitian. To circumvent this problem, we use the ideas of
Ref. \cite{silveirinha_chern_2015, morgado_analytical_2015} to
describe the material dispersion in terms of additional variables.
Specifically, to obtain an Hermitian eigenvalue problem, the system
\eqref{E:Maxwell_equation_time} may be transformed into the
equivalent generalized problem \cite{silveirinha_chern_2015}
\begin{align} \label{E:Maxwell_equation_generalized}
  \op{L} \cdot \vec{Q} = i \cdot \frac{\partial}{\partial t} \vec{M}_g \cdot \vec{Q} + i\vec{J}_g,
\end{align}
where $\op{L}$ and $\vec{M}_g$ are (frequency independent) operators
that can be constructed as explained in Ref.
\cite{silveirinha_chern_2015} and $\vec{Q}=\left(\vec{F} \quad
\vec{Q}_1 \quad \vec{Q}_2 \quad \dots \right)^T$ is a generalized
state vector whose first six-components determine the
electromagnetic field $\vec{F}$. This transformation is possible
provided all the materials are lossless. The number of additional
variables $\vec{Q}_i$'s is determined by the number of poles of
$\vec{M}(\omega)$. In general both $\op{L}$ and $\vec{M}_g$ are
space dependent. The generalized excitation vector is of the form
$\vec{J}_g = \left(\vec{J} \quad \vec{0} \quad \vec{0} \quad \dots
\right)^T$ \cite{silveirinha_chern_2015, morgado_analytical_2015}.

The eigenmodes of the generalized system
\eqref{E:Maxwell_equation_generalized} satisfy:
\begin{equation}  \label{E:eigenmodes_generalized_Maxwell}
 \vec{M}_g^{-1 } \cdot \op{L} ~\vec{Q}_{n\k} = \omega_{n\k} \vec{Q}_{n\k}
\end{equation}
It can be demonstrated that $\vec{M}_g^{-1} \cdot \op{L}$ is a
Hermitian operator with respect to the weighted inner product
\begin{equation} \label{E:weighted_inner_product}
 \braket{\vec{Q}_A|\vec{Q}_B}= \frac{1}{2} \int d^3\vec{r}~ \vec{Q}_A^\ast(\vec{r}) \cdot \vec{M}_g(\vec{r}) \cdot \vec{Q}_B(\vec{r}).
\end{equation}
Then, according to the spectral theorem, the eigenfunctions
$\vec{Q}_{n\k}$ define a complete set of basis vectors and can be
normalized as
\begin{equation}  \label{E:orthonormality_generalized_eigenmodes}
 \braket{\vec{Q}_{m\vec{q}}|\vec{Q}_{n\k}}= \delta_{n,m}\delta_{\vec{q},\k}.
\end{equation}
As demonstrated in Ref.\cite{silveirinha_chern_2015} the above
normalization condition implies that the electromagnetic field
sub-component of the state vector is normalized as in Eq.
\eqref{E:relation_normalization_modes}.

Next, we introduce a Green's function $ \db{G}_Q$ for the
generalized problem defined as the solution of:
\begin{align} \label{E:def_spectral_green_dispersive}
\left( \op{L} -  \omega \vec{M}_g \right) \cdot \db{G}_Q(
\vec{r},\vec{r}_1) = i\delta(\vec{r}-\vec{r}_1) \vec{1}.
\end{align}
Evidently, the tensor $\db{G}_Q$ can be expanded as
$\db{G}_Q(\vec{r},\vec{r}_1)= \sum_{n\k}   \vec{Q}_{n\k}(\vec{r})
\otimes \boldsymbol{\alpha}_{n\k}$ where $\boldsymbol{\alpha}_{n\k}$
are unknown vectorial coefficients
 \cite{sakoda_optical_2004,novotny_principles_2012}.
Using the completeness of the eigenfunctions and standard ideas
\cite{sakoda_optical_2004}, it is readily found that:
\begin{equation}
 \boldsymbol{\alpha}_{n\k}    =  \frac{ i }{2\left( \omega_{n\k} - \omega \right) }
 \vec{Q}_{n\k}^\ast(\vec{r}_1).
\end{equation}
Thus, it follows that the generalized Green's function has the
eigenmode expansion:
\begin{equation}
 \db{G}_Q(\vec{r},\vec{r}_1)= \sum_{n\k}  \frac{ i }{2\left( \omega_{n\k} - \omega \right) }  \vec{Q}_{n\k}(\vec{r}) \otimes   \vec{Q}_{n\k}^\ast(\vec{r}_1).
\end{equation}
The Green's function of the original system
\eqref{E:Maxwell_Green's_function_freq} is the restriction of
$\db{G}_Q$ to its first six by six components:
\begin{equation} \label{E:green_function_spectral}
 \db{G}(\vec{r},\vec{r}_1, \omega)= \sum_{n\k}  \frac{ i }{2\left( \omega_{n\k}   - \omega \right) }   \vec{F}_{n\k}(\vec{r}) \otimes
 \vec{F}_{n\k}^\ast(\vec{r}_1),
\end{equation}
with $\vec{F}_{n\k}(\vec{r})$ normalized as in Eq.
\eqref{E:relation_normalization_modes}.



\bibliographystyle{ieeetr}

\bibliography{Biblio}

\end{document}